\documentclass[graybox, secnum]{svmult}

\usepackage{mathptmx}       
\usepackage{helvet}         
\usepackage{courier}        
\usepackage{type1cm}        
\usepackage{makeidx}         
\usepackage{graphicx}        
\usepackage{multicol}        
\usepackage[bottom]{footmisc}
\usepackage{hyperref}        
\usepackage{soul}            
\hypersetup{colorlinks=true,urlcolor=blue}

\newcommand{\ft}{pdf}        

\newcommand{\makeSymbol}[1]{\mathord{\vcenter{\hbox{#1}}}}
\newcommand{\Symbol}[1]{\makeSymbol{\includegraphics[scale=0.5]{#1.\ft}}}

\usepackage{amsmath}
\usepackage{amsfonts}
\usepackage{amssymb}
\usepackage{bbm}
\usepackage[makeroom]{cancel}
\usepackage{cite}
\usepackage{color}
\usepackage{enumerate}
\usepackage{epsfig}
\usepackage{listings}
\usepackage{soul}
\usepackage{url}
\usepackage{xspace}

\newcommand{\bra}[1]{\langle #1 |}
\newcommand{\ket}[1]{|#1\rangle}
\newcommand{\braket}[2]{\langle #1 | #2 \rangle}
\newcommand{\bbraket}[2]{\bigl\langle #1 \big| #2 \bigr\rangle}
\newcommand{\bbra}[1]{\bigl\langle #1 \bigr|}
\newcommand{\bket}[1]{\bigl|#1\bigr\rangle}

\newcommand{\eg}{e.g.\@\xspace}
\newcommand{\ie}{i.e.\@\xspace}
\newcommand{\Eq}[1]{Eq.\@\xspace\eqref{#1}}
\newcommand{\Eqs}[1]{Eqs.\@\xspace\eqref{#1}}

\newcommand{\updown}[2]{^{#1}_{\phantom{#1}#2}}
\newcommand{\downup}[2]{_{#1}^{\phantom{#1}#2}}

\newcommand{\Id}{\mathbbm{1}}
\newcommand{\R}{\mathbbm{R}}
\newcommand{\C}{\mathbbm{C}}

\newcommand{\idx}[1]{#1}

\newcommand{\DD}[4]{{D^{(#1)#2}}_{#3}(#4)}
\newcommand{\CG}[4]{C^{(#1#2)}{}\downup{#3}{#4}}
\newcommand{\CGi}[4]{C^{(#1#2)}{}\updown{#3}{#4}}
\newcommand{\Tau}[4]{({\tau^{(#1)}_{#2}}){}\updown{#3}{#4}}
\newcommand{\threej}[6]{\begin{pmatrix} #1&#2&#3 \\ #4&#5&#6 \end{pmatrix}}
\newcommand{\sixj}[6]{\begin{Bmatrix} #1&#2&#3 \\ #4&#5&#6 \end{Bmatrix}}

\usepackage[square,numbers]{natbib}
\makeindex             

\begin{document}
\title*{Graphical Calculus of Spin Networks}
\author{Emanuele Alesci\thanks{corresponding author}, Ilkka Mäkinen and Jinsong Yang}
\institute{Emanuele Alesci \at Institute for Theoretical Physics \& Cosmology, Zhejiang University of Technology, Hangzhou, 310023, China \\
United Center for Gravitational Wave Physics (UCGWP), Zhejiang University of Technology, Hangzhou, 310023, China \\
\email{emanuele.alesci@gmail.com}
\and Ilkka Mäkinen \at National Centre for Nuclear Research, Pasteura 7, 02-093 Warsaw, Poland \\
Faculty of Physics, University of Warsaw, Pasteura 5, 02-093 Warsaw, Poland \\
\email{ilkka.makinen@ncbj.gov.pl}
\and Jinsong Yang \at School of Physics, Guizhou University, Guiyang 550025, China \\
\email{jsyang@gzu.edu.cn}}
%
%
\maketitle


\abstract{Graphical techniques provide a very useful practical device for calculations involving the so-called spin network states, which encode the quantum degrees of freedom of spatial geometry in loop quantum gravity. Graphical calculus of $SU(2)$, which has been originally introduced in the literature in order to deal with calculations arising from the coupling of angular momenta in quantum mechanics, can be used as a simple but powerful method for computing the action of various physically interesting operators in the spin network representation. Compared with conventional manipulation of algebraic expressions, calculations in the graphical approach are typically more convenient, concise and visually transparent. The goal of this chapter is to provide an accessible introduction to graphical methods in $SU(2)$ recoupling theory and a brief description of their use as a tool for practical calculations in loop quantum gravity. We introduce the basics of the graphical formalism, and establish the representation of the elementary states and operators of loop quantum gravity in graphical form. Several example calculations are given to illustrate the use of the graphical techniques, including a computation of the matrix elements of a Hamiltonian constraint operator in the spin network basis.
}

\section*{Keywords} 


Loop quantum gravity; Graphical calculus; Spin network states; Intertwiners; Hamiltonian constraint operator; $SU(2)$ representation theory; $SU(2)$ recoupling theory; Angular momentum diagrams

\section{Introduction}

Loop quantum gravity provides a non-perturbative and background-independent approach to the quantization of general relativity (see \eg \cite{Ashtekar:2017,Rovelli:2004tv,Rovelli:2014,Thiemann:2007pyv} for books and \cite{Thiemann:2002nj,Ashtekar:2004eh,Han:2005km,Giesel:2012ws,Baez:1999sr,Rovelli:2011eq,Perez:2012wv} for review articles). The theory incorporates a key lesson of general relativity -- that the gravitational field and the geometry of spacetime are essentially the same physical entity -- and is built on a rigorous mathematical foundation. In the past thirty years, loop quantum gravity has developed into one of the main candidates for a quantum theory of gravitation, both its canonical (Hamiltonian) and covariant (Lagrangian) formulations.

At the heart of loop quantum gravity are the so-called spin network states, which encode the quantum degrees of freedom of discrete, quantized spatial geometries \cite{Rovelli:1994ge, Rovelli:1995}. The use of graphical techniques for calculations involving spin network states has a long history in loop quantum gravity. The articles introducing the spin network representation \cite{Rovelli:1994ge, Rovelli:1995} and performing some of the first detailed computations with operators in the spin network basis (\eg \cite{Borissov:1997ji, DePietri:1996tvo}) made extensive use of graphical methods which can be traced back to Penrose's diagrammatic tensor calculus \cite{Penrose} and the tangle-theoretic recoupling theory due to Kauffman and Lins \cite{Kauffman}. A related but arguably more powerful graphical formalism for $SU(2)$ recoupling theory was originally developed to deal with calculations arising from the coupling of angular momenta in quantum mechanics, and has been presented in slightly different versions by Yutsis et.~al. \cite{Yutsis:1962bk}, Brink and Satchler \cite{brink1968angular}, and Varshalovich et.~al. \cite{Varshalovich:1988ye}. All these methods consist of two ingredients: graphical representation and graphical calculation. An algebraic formula involving objects of $SU(2)$ recoupling theory is first represented by a corresponding graphical diagram in a unique and unambiguous way. Then the graphical calculation will be performed following certain simple rules for transforming graphical expressions, which correspond uniquely to algebraic manipulations of the corresponding non-graphical formulae.

The graphical methods originating from the literature of quantum angular momentum provide a very useful tool for practical calculations in loop quantum gravity. Examples of calculations to which $SU(2)$ graphical calculus has been successfully applied include studying the action of geometric operators (such as the volume operator \cite{Rovelli:1994ge,Ashtekar:1997fb}) \cite{DePietri:1996tvo, Bianchi:2008es, Yang:2015wka,Yang:2016kia, Yang:2019xms} and the Hamiltonian constraint operator \cite{Alesci:2010gb, Alesci:2013kpa,Gaul:2000ba,Alesci:2015wla,Makinen:2019} in the spin network basis; examining the consistency \cite{Yang:2021den, Alesci:2011ia,Thiemann:2013lka} between the dynamics defined by Thiemann's Hamiltonian constraint \cite{Thiemann:1996aw} and the EPRL (Engle--Pereira--Rovelli--Livine) spin foam model \cite{Engle:2007wy,Freidel:2007py,Kaminski:2009fm}; analysis of the semiclassical limit of the Barrett--Crane spin foam model \cite{Alesci:2007tx}, and a proof of invariance of a lattice formulation of $BF$ theory under Pachner moves \cite{Kawamoto:1999jv}. Pedagogical presentations of these graphical techniques and their use in loop quantum gravity have been given in \cite{Martin-Dussaud:2019ypf,Makinen:2019rou}.

Various generalizations and extensions of the graphical method have been developed to extend its reach beyond calculations involving purely $SU(2)$ recoupling theory. In particular, graphical calculus of the group $SL(2,\C)$ plays an essential role in the analysis of EPRL spin foam amplitudes (see \eg \cite{Perez:2012wv} for a review, or the recent articles \cite{Dona:2020,Dona:2022,Frisoni:2022} for a selection of examples). The graphical approach has also been extended to $SU(3)$ in connection with Yang--Mills theory coupled to loop quantum gravity \cite{Liegener:2016mgc}, to $OSp(1|2)$ in the context of loop quantum supergravity \cite{Ling:1999}, and to the quantum group $SU(2)_q$ (see \eg \cite{Borissov:1995, Dittrich:2016}). Furthermore, graphical techniques been applied to calculations in quantum-reduced loop gravity \cite{Alesci:2013xd,Alesci:2014c}.

Such advanced applications, however, are decidedly outside the scope of our discussion. The purpose of this chapter is limited to giving an accessible introduction to the graphical calculus of $SU(2)$ recoupling theory and explaining how these graphical techniques can be applied to concrete calculations in loop quantum gravity in the spin network representation. After introducing the elements of the graphical formalism in Sec.~\ref{sec:elements} and the highly useful fundamental theorem of graphical calculus in Sec.~\ref{sec:calculating}, we establish the graphical representation of the elementary states and operators of loop quantum gravity in Sec.~\ref{sec:LQG} and illustrate their use with an example calculation of the matrix elements of the Hamiltonian constraint operator. The chapter is concluded with a brief summary in Sec.~\ref{sec:summary}.

\section{Elements of graphical calculus}
\label{sec:elements}

In this section we introduce a diagrammatic notation for the fundamental objects of $SU(2)$ representation theory and establish the basic rules for manipulating such diagrams. The conventions we choose are not in complete agreement with any standard reference on the subject, although they follow quite closely the conventions of Brink and Satchler \cite{brink1968angular}. Our discussion is interlaced with a concise summary of the necessary elements of $SU(2)$ representation theory, but does not provide a comprehensive review. If necessary, we encourage the reader to consult some of the many available references on Lie groups and the quantum theory of angular momentum (\eg \cite{brink1968angular, Edmonds, Fecko, Wigner-bk}).

\subsection{$SU(2)$ representation theory}

The fundamental representation of $SU(2)$, the group of unitary $2\times 2$ matrices with determinant $+1$, is realized by the natural action of these matrices on $\C^2$. Higher irreducible representations can be described in terms of the states $\ket{jm}$, which are the familiar eigenstates of angular momentum ($\ket{jm}$ is an eigenstate of the operators $J^2$ and $J_z$ with respective eigenvalues $j(j+1)$ and $m$). The spin-$j$ representation of $SU(2)$ is realized on the $(2j+1)$-dimensional space ${\cal H}_j$ spanned by the states $\ket{jm}$ with a fixed value of the quantum number $j$. 

The matrices representing elements of $SU(2)$ on ${\cal H}_j$ are known as the \idx{Wigner matrices}. We denote their matrix elements by
\begin{equation}
	\DD{j}{m}{n}{g} = \bra{jm}D^{(j)}(g)\ket{jn}.
	\label{}
\end{equation}
To express these matrix elements in graphical form, we adopt the notation
\begin{equation}
	\DD{j}{m}{n}{g} \; = \; \Symbol{Dmn}\vspace{-12pt}
	\label{g:D}
\end{equation}
with the tip of the triangle pointing towards the upper index.

A fundamental invariant tensor on ${\cal H}_j$ (the \idx{epsilon tensor}) is defined by
\begin{equation}
	\epsilon^{(j)}_{mn} = (-1)^{j-m}\delta_{m,-n}, \qquad \epsilon_{\phantom{m}}^{(j)mn} = (-1)^{j-m}\delta_{m,-n}.
	\label{eps}
\end{equation}
Note that we define the tensor with upper indices to be numerically equal to its counterpart with lower indices. From its definition, we see that the epsilon tensor satisfies the properties
\begin{equation}
	\epsilon^{(j)}_{nm} = (-1)^{2j}\epsilon^{(j)}_{mn}, \qquad \epsilon^{(j)}_{mm'}\epsilon_{\phantom{m'}}^{(j)m'n} = (-1)^{2j}\delta_m^n.
	\label{eps-properties}
\end{equation}
The epsilon tensor can be used to raise and lower indices of vectors on ${\cal H}_j$. We choose the convention
\begin{equation}
	v^m = \epsilon_{\phantom{m}}^{(j)mn}v_n, \qquad v_m = v^n\epsilon^{(j)}_{nm}.
	\label{}
\end{equation}

The tensor $\epsilon^{(j)}_{mn}$ is represented graphically by a line carrying an arrow and a label indicating the representation $j$. The two ends of the line correspond to the indices $m$ and $n$, with the arrow pointing from $m$ to $n$:
\begin{equation}
	\epsilon^{(j)}_{mn}\; = \;\;\Symbol{epsilon-indices} \quad . \vspace{-12pt}
	\label{g:eps}
\end{equation}
The tensor $\epsilon^{(j)mn}_{\phantom{m}}$, which is numerically equal to $\epsilon^{(j)}_{mn}$, is represented by the same diagram. We also introduce a line without an arrow to represent the unit tensor (Kronecker delta):
\begin{equation}
	\delta^m_n\; = \;\;\Symbol{delta-indices} \quad . \vspace{-12pt}
	\label{g:del}
\end{equation}

The properties \eqref{eps-properties} translate to identities satisfied by the graphical objects \eqref{g:eps} and \eqref{g:del}. The first property implies that the direction of the arrow can be reversed according to the rule
\begin{equation}
	\Symbol{invepsilon}\quad = \quad(-1)^{2j}\;\Symbol{epsilon} \quad . \vspace{-12pt}
	\label{g:eps_nm}
\end{equation}
In the graphical formalism, contraction of magnetic indices is represented by connecting the two ends of lines corresponding to the contracted index. The second property \eqref{eps-properties} then shows that two consecutive arrows behave according to the rules
\begin{align}
	\Symbol{epseps}\quad &= \quad(-1)^{2j}\;\Symbol{delta} \quad ,  \label{g:epseps} \\[-2ex]
	\Symbol{epseps2}\quad &= \quad\Symbol{delta} \quad , \label{g:epseps2}
\end{align}
where the second relation is obtained by combining \Eqs{g:eps_nm} and \eqref{g:epseps}.

If the names of the magnetic indices are irrelevant or clear from the context, it is typical to not write the indices explicitly in graphical diagrams such as \eqref{g:eps} or \eqref{g:del}, as we have already done in \Eqs{g:eps_nm}--\eqref{g:epseps2}.

Using the fact that the epsilon tensor is invariant under the action of $SU(2)$ on its indices, one can deduce the expression
\begin{equation}
	\DD{j}{m}{n}{g^{-1}} = \epsilon_{\phantom{m'}}^{(j)mm'}\epsilon^{(j)}_{nn'}\DD{j}{n'}{m'}{g}
	\label{D-inv}
\end{equation}
for the matrix elements of the inverse matrix $D^{(j)}(g^{-1})$ in terms of the matrix $D^{(j)}(g)$. Equivalently, we have the graphical equation
\begin{equation}
	\Symbol{Dginv} \quad = \quad \Symbol{epsDeps} \quad . 
	\label{}
\end{equation}
The great orthogonality theorem for compact groups states that the matrix elements of the Wigner matrices satisfy
\begin{equation}
	\int dg\,\overline{\displaystyle \DD{j}{m}{n}{g}} \DD{j'}{m'}{n'}{g} = \frac{1}{d_j}\delta_{jj'}\delta_m^{m'}\delta^n_{n'},
	\label{D-orth}
\end{equation}
where $dg$ is the Haar measure of $SU(2)$, and $d_j = 2j+1$ is a common abbreviation for the dimension of the space ${\cal H}_j$. To express the orthogonality theorem in graphical form, it is convenient to start by using \Eq{D-inv} together with unitarity, \ie $D^{(j)}(g^\dagger) = D^{(j)}(g^{-1})$, to derive an equation in which the complex conjugate has been removed. Then expressing this equation in graphical notation, one obtains
\begin{equation}
	\int dg\quad \Symbol{DD-vertical} \quad = \quad \frac{1}{d_j}\delta_{jj'}\;\Symbol{twoeps} \quad . 
	\label{g:intDD}
\end{equation}

\subsection{Wigner $3j$-symbol}
\label{sec:3j}

Two sets of basis states on the tensor product space ${\cal H}_{j_1}\otimes {\cal H}_{j_2}$ are given by the ``uncoupled'' states $\ket{j_1m_1}\ket{j_2m_2}$ and the ``coupled'' states $\ket{j_1j_2;jm}$, which are defined as the eigenstates of the complete sets of mutually commuting operators $\bigl\{(J^{(1)})^2, J^{(1)}_z, (J^{(2)})^2, J^{(2)}_z\bigr\}$ and $\bigl\{(J^{(1)})^2, (J^{(2)})^2, (J^{(1)} + J^{(2)})^2, J^{(1)}_z + J^{(2)}_z\bigr\}$. The change of basis between the two bases is given by
\begin{equation}
	\ket{j_1m_1}\ket{j_2m_2} = \sum_{jm} \CG{j_1j_2}{j}{m_1m_2}{m}\ket{j_1j_2;jm},
	\label{uncoupled}
\end{equation}
where $\CG{j_1j_2}{j}{m_1m_2}{m}$ are the \idx{Clebsch--Gordan coefficients} of $SU(2)$. In the physics literature, the Clebsch--Gordan coefficients are usually denoted by a notation such as $\braket{j_1m_1,j_2m_2}{jm}$. The purpose of our notation is to indicate the index structure of the Clebsch--Gordan coefficient when seen as an $SU(2)$ tensor. The phases of the Clebsch--Gordan coefficients are fixed by the nearly universally followed Condon--Shortley convention (see \eg \cite{brink1968angular, Edmonds, Varshalovich:1988ye}). Under the Condon--Shortley convention, all Clebsch--Gordan coefficients are real-valued, and the coefficients $\CGi{j_1j_2}{j}{m_1m_2}{m}$ of the inverse transformation, expressing the coupled states $\ket{j_1j_2;jm}$ in the uncoupled basis, are numerically equal to the coefficients $\CG{j_1j_2}{j}{m_1m_2}{m}$ themselves. The Clebsch--Gordan coefficients also feature in the so-called Clebsch--Gordan series
\begin{equation}
	\DD{j_1}{m_1}{n_1}{g}\DD{j_2}{m_2}{n_2}{g} = \sum_{jmn} \CGi{j_1j_2}{j}{m_1m_2}{m}\CG{j_1j_2}{j}{n_1n_2}{n}\DD{j}{m}{n}{g},
	\label{DD}
\end{equation}
which can be derived by studying how the state \eqref{uncoupled} transforms under an $SU(2)$ rotation.

An object closely related to the Clebsch--Gordan coefficient is the Wigner \idx{$3j$-symbol}. The $3j$-symbol is constructed by using the epsilon tensor to lower the last index of the Clebsch--Gordan coefficient (and multiplying with a numerical factor):
\begin{align}
	\threej{j_1}{j_2}{j_3}{m_1}{m_2}{m_3} &= \frac{1}{\sqrt{d_{j_3}}}(-1)^{j_1-j_2+j_3}\CG{j_1j_2}{j_3}{m_1m_2}{n}\epsilon^{(j_3)}_{nm_3}. 
	\label{3j}
\end{align}
The value of the $3j$-symbol is nonvanishing only if the spins $j_1$, $j_2$ and $j_3$ fulfill the triangular conditions, or Clebsch--Gordan conditions
\begin{equation}
	|j_1 - j_2| \leq j_3 \leq j_1 + j_2, \qquad \text{and} \qquad j_1 + j_2 + j_3 \text{ is an integer},
	\label{}
\end{equation}
and the magnetic quantum numbers sum up to zero:
\begin{equation}
	m_1 + m_2 + m_3 = 0.
	\label{}
\end{equation}
In the graphical notation, the $3j$-symbol is taken as a fundamental object due to its higher degree of symmetry over the Clebsch--Gordan coefficient. The $3j$-symbol is represented graphically by a node with three lines connected to it:
\begin{equation}
	\threej{j_1}{j_2}{j_3}{m_1}{m_2}{m_3}\; = \;\Symbol{3j} \quad .
	\label{g:3j}
\end{equation}
The cyclic order of the spins is indicated by a $+$ or $-$ sign next to the node. The two signs represent respectively anticlockwise and clockwise ordering of the spins. Thus, a node with a minus sign represents the $3j$-symbol
\begin{equation}
	\Symbol{3j-minus} \quad = \quad \threej{j_1}{j_3}{j_2}{m_1}{m_3}{m_2} \quad = \quad \Symbol{3j-132} \quad .
	\label{}
\end{equation}
From \Eqs{3j} and \eqref{g:3j} it follows that the graphical representation of the Clebsch--Gordan coefficient is given by
\begin{equation}
	\CG{j_1j_2}{j}{m_1m_2}{m} \; = \; (-1)^{j_1-j_2-j}\sqrt{d_j}\;\;\Symbol{clebsch} \quad . 
	\label{}
\end{equation}
The Clebsch--Gordan series \eqref{DD} then takes the graphical form
\begin{equation}
	\Symbol{DD-clebsch} \quad = \quad \sum_j d_j\;\Symbol{CDC} \quad .
	\label{g:DD}
\end{equation}

The $3j$-symbol enjoys a number of convenient symmetry properties. Interchanging any two columns in the symbol is equivalent to multiplying the symbol by $(-1)^{j_1+j_2+j_3}$ (so in particular, the symbol is invariant under cyclic permutations of the columns). The same multiplicative factor results if the sign of all the magnetic numbers is reversed. Translated to graphical notation, these statements imply that the diagram representing the $3j$-symbol satisfies the following basic properties:
\begin{equation}
	\Symbol{3j-minus} \quad = \quad (-1)^{j_1+j_2+j_3}\Symbol{3j} \quad ,
	\label{g:3j-minus}
\end{equation}
\ie reversing the sign is equivalent to multiplying the diagram with the phase factor $(-1)^{j_1+j_2+j_3}$, and
\begin{equation}
	\Symbol{3j-arrows} \quad = \quad \Symbol{3j} \quad ,
	\label{g:3j-arrows}
\end{equation}
\ie a node with an identically oriented arrow on each line is equivalent to a node with no arrows. The $3j$-symbol also satisfies certain orthogonality relations, which encode the orthonormality of the coupled and uncoupled bases, and which can be expressed graphically as
\begin{align}
	\Symbol{3j-orth1} \quad &= \quad\delta_{jj'}\frac{1}{d_j}\;\Symbol{delta} \quad , \label{g:3j-orth-mm} \\[1ex]
	\sum_j d_j\Symbol{3j-orth2} \quad &= \quad \Symbol{3j-deltadelta} \quad . \label{g:3j-orth-jm}
\end{align}
Further properties of the $3j$-symbol include the identity
\begin{equation}
	\Symbol{3j-zero} \quad = \quad \delta_{jj'}	\frac{1}{\sqrt{d_j}}\;\Symbol{eps-vertical} \quad ,
	\label{}
\end{equation}
which shows how the $3j$-symbol reduces to the epsilon tensor when one of the spins is equal to zero.

Certain familiar objects of $SU(2)$ representation theory can be expressed in graphical form by relating them to the $3j$-symbol. The anti-Hermitian generators $\tau_i^{(j)}$ in the spin-$j$ representation are defined by the matrix elements
\begin{equation}
	\Tau{j}{i}{m}{n} = -i\bra{jm}J_i\ket{jn}.
	\label{}
\end{equation}
However, the Wigner--Eckart theorem (see \eg \cite{brink1968angular, Edmonds, Varshalovich:1988ye}) states that the matrix elements of the angular momentum operator are given by
\begin{equation}
	\bra{jm}J_i\ket{jn} = (j||\vec J\,||j)\CG{j1}{j}{ni}{m},
	\label{wigner-eckart}
\end{equation}
where the so-called reduced matrix element $(j||\vec J\,||j)$ is independent of the magnetic indices $m$, $n$ and $i$. The value of the reduced matrix element, $(j||\vec J\,||j) = \sqrt{j(j+1)}$, can  be deduced \eg by considering a complete contraction of \Eq{wigner-eckart} with itself, and recalling that the Condon--Shortley convention fixes the sign of the coefficient $\CG{j1}{j}{j0}{j}$ to be positive. In this way one finds that the generators are represented graphically by
\begin{equation}
	\Tau{j}{i}{m}{n} \;\; = \;\; iW_j\;\;\Symbol{tau} \quad , 
	\label{g:tau}
\end{equation}
where we have introduced the abbreviation
\begin{equation}
	W_j = \sqrt{j(j+1)(2j+1)}.
	\label{}
\end{equation}
Moreover, if we use the diagram
\begin{equation}
	v^m \; = \; \Symbol{vector}
	\label{}
\end{equation}
to denote a vector with an index in the $j=1$ representation, we can write down the expression
\begin{equation}
	\epsilon_{ijk}u^iv^jw^k \; = \;i\sqrt{6}\;\Symbol{triple} \quad
	\label{g:triple}
\end{equation}
for the triple product of three vectors contracted with the completely antisymmetric tensor $\epsilon_{ijk}$. This is based on the observation that the $3j$-symbol with spins $j_1=j_2=j_3=1$, being antisymmetric under the interchange of any two columns, must be proportional to $\epsilon_{m_1m_2m_3}$, and that a triple product expressed in the Cartesian basis is related to the same product calculated in the spherical basis\footnote{The spherical components of a vector $\vec v$ in $\R^3$ are given in terms of the Cartesian components by
\[
	v^+ = -\frac{1}{\sqrt 2}(v^x - iv^y), \qquad v^0 = v^z, \qquad v^- = \frac{1}{\sqrt 2}(v^x+iv^y).
\]
} by $\epsilon_{ijk}u^iv^jw^k = -i\epsilon_{m_1m_2m_3}u^{m_1}v^{m_2}w^{m_3}$.

\subsection{Invariant tensors. Recoupling theory}
\label{sec:intertwiners}

In the context of loop quantum gravity, a particularly important property of the $3j$-symbol is its invariance under $SU(2)$ transformations. With the help of \Eq{DD}, one can show that
\begin{equation}
	\DD{j_1}{m_1}{n_1}{g}\DD{j_2}{m_2}{n_2}{g}\DD{j_3}{m_3}{n_3}{g} \threej{j_1}{j_2}{j_3}{m_1}{m_2}{m_3} = \threej{j_1}{j_2}{j_3}{n_1}{n_2}{n_3}.
	\label{}
\end{equation}
A notation such as
\begin{equation}
	\iota^{(j_1j_2j_3)}_{m_1m_2m_3}\; = \;\Symbol{3j}
	\label{g:iota3}
\end{equation}
is commonly adopted to denote the $3j$-symbol if one wishes to emphasize its interpretation as an invariant tensor. The indices of the tensor can be raised with the epsilon tensor in the usual way. In particular, the relation \eqref{g:3j-arrows} shows that the tensor $\bigl(\iota^{(j_1j_2j_3)}\bigr)^{m_1m_2m_3}$, obtained by raising all the indices of the tensor \eqref{g:iota3}, is numerically equal to $\iota^{(j_1j_2j_3)}_{m_1m_2m_3}$. Furthermore, the orthogonality relation \eqref{g:3j-orth-mm} implies that the tensor \eqref{g:iota3} is normalized to 1.

The $3j$-symbol and the epsilon tensor are the basic building blocks out of which invariant tensors of higher valence can be constructed. For example, using the epsilon tensor to contract two $3j$-symbols on one index, we obtain the four-valent invariant tensor
\vspace{-12pt}
\begin{align}
	\bigl(\iota^{(j_1\cdots j_4)}_k\bigr)_{m_1\cdots m_4} \; = \; \Symbol{iota412} \quad .
	\label{g:iota4}
\end{align}
From the invariance of the $3j$-symbol and the epsilon tensor it follows that the tensor \eqref{g:iota4} is also invariant under the action of $SU(2)$. Note that invariant tensors whose index structure consists of a mixture of upper and lower indices satisfy $SU(2)$ invariance in the form where a representation matrix acts on each lower index while the inverse matrix acts on each upper index. For instance, the tensor \eqref{g:iota4} with the first index raised satisfies
\begin{equation}
\DD{j_1}{n_1}{m_1}{g^{-1}}\DD{j_2}{m_2}{n_2}{g}\cdots\DD{j_4}{m_4}{n_4}{g} \bigl(\iota^{(j_1\cdots j_4)}_k\bigr)\updown{m_1}{m_2m_3m_4} = \bigl(\iota^{(j_1\cdots j_4)}_k\bigr)\updown{n_1}{n_2n_3n_4}.
	\label{}
\end{equation}
In loop quantum gravity, invariant tensors such as \eqref{g:iota3} and \eqref{g:iota4} are typically called \idx{intertwiners}.

The tensors \eqref{g:iota4} corresponding to all values of the internal spin $k$ allowed by the triangular conditions span the intertwiner space ${\rm Inv}\,\bigl({\cal H}_{j_1}\otimes\cdots\otimes{\cal H}_{j_4}\bigr)$. However, unlike the three-valent intertwiner \eqref{g:iota3}, the four-valent intertwiners \eqref{g:iota4} are not normalized; a short calculation using \Eq{g:3j-orth-mm} shows that
\begin{equation}
	\bbraket{\iota^{(j_1\cdots j_4)}_k}{\iota^{(j_1\cdots j_4)}_l} \; = \;\; \Symbol{iota4-product} \;\; = \; \frac{1}{d_k}\delta_{kl}.
	\label{iota4-norm}
\end{equation}
Hence, an orthonormal basis on ${\rm Inv}\,\bigl({\cal H}_{j_1}\otimes\cdots\otimes{\cal H}_{j_4}\bigr)$ is given by the intertwiners
\vspace{-4pt}
\begin{equation}
	\bigl(\nu^{(j_1\cdots j_4)}_k\bigr)_{m_1m_2m_3m_4} = \sqrt{d_k}\bigl(\iota^{(j_1\cdots j_4)}_k\bigr)_{m_1m_2m_3m_4}.
	\label{}
\end{equation}
The process of using epsilon tensors to connect $3j$-symbols to each other can be continued to construct intertwiners of arbitrarily high valence. An $N$-valent intertwiner obtained in this way has the form
\begin{equation}
	\iota^{(k_1\cdots k_{N-3})}_{m_1\cdots m_N} \; = \quad \Symbol{iotaN} \quad . \vspace{-16pt}
	\label{g:iotaN}
\end{equation}
where the label $\vec k$ denotes the collection of internal spins $k_1, \dots, k_{N-3}$. As the internal spins range over all their possible values, the intertwiners \eqref{g:iotaN} span\footnote{
Seen as a state on ${\cal H}_{j_1}\otimes\cdots\otimes{\cal H}_{j_N}$, the tensor \eqref{g:iotaN} is an eigenstate of the operators
\[
	\bigl(J^{(1)} + J^{(2)}\bigr)^2, \qquad \bigl(J^{(1)} + J^{(2)} + J^{(3)}\bigr)^2, \qquad \dots, \qquad \bigl(J^{(1)} + J^{(2)} + \dots + J^{(N-1)}\bigr)^2
\]
with eigenvalues determined by the spins $k_1, \dots, k_{N-3}$, and of the operators
\[
	\bigl(J^{(1)} + J^{(2)} + \dots + J^{(N)}\bigr)^2, \qquad J^{(1)}_z + J^{(2)}_z + \dots + J^{(N)}_z
\]
with vanishing eigenvalues. Since these operators form a complete set of commuting operators on ${\cal H}_{j_1}\otimes\cdots\otimes{\cal H}_{j_N}$, and the invariant subspace ${\rm Inv}\,\bigl({\cal H}_{j_1}\otimes\cdots\otimes{\cal H}_{j_N}\bigr)$ is characterized by eigenvalues $j=0$ and $m=0$ of the total angular momentum, it follows that the states \eqref{g:iotaN} provide a complete basis on ${\rm Inv}\,\bigl({\cal H}_{j_1}\otimes\cdots\otimes{\cal H}_{j_N}\bigr)$.
}
the $N$-valent intertwiner space ${\rm Inv}\,\bigl({\cal H}_{j_1}\otimes\cdots\otimes{\cal H}_{j_N}\bigr)$. As before, this basis is orthogonal but not normalized; an orthonormal basis is given by the intertwiners
\begin{equation}
	\bigl(\nu^{(j_1\cdots j_N)}_{\vec k}\bigr)_{m_1\cdots m_N} = \sqrt{d_{k_1}\cdots d_{k_{N-3}}}\;\bigl(\iota^{(j_1\cdots j_N)}_{\vec k}\bigr)_{m_1\cdots m_N}.
	\label{}
\end{equation}

For intertwiners of valence greater than three, there generally exist many different ways of coupling the spins $j_1, \dots, j_N$ to the internal spins of the intertwiner, giving rise to several inequivalent bases of the corresponding intertwiner space. For example, the intertwiners
\vspace{-8pt}
\begin{equation}
	\bigl(\widetilde\iota^{\,(j_1\cdots j_4)}_l\bigr)_{m_1\cdots m_4} \; = \; \Symbol{iota413} \quad ,
	\label{g:iota4-13}
\end{equation}
where the first and third spins are now coupled to the internal spin, provide another basis of the space ${\rm Inv}\,\bigl({\cal H}_{j_1}\otimes\cdots\otimes{\cal H}_{j_4}\bigr)$. The change of basis between the bases \eqref{g:iota4} and \eqref{g:iota4-13} is given by
\begin{equation}
	\Symbol{iota413} \quad = \quad \sum_k d_k(-1)^{j_2+j_3+k+l}\sixj{j_1}{j_2}{k}{j_4}{j_3}{l}\;\Symbol{iota412} \quad . 
	\label{g:13-in-basis-12}
\end{equation}
Here the object denoted by the curly brackets is a Wigner \idx{$6j$-symbol}. In graphical notation, the $6j$-symbol can be defined as the diagram
\begin{equation}
	\sixj{j_1}{j_2}{j_3}{k_1}{k_2}{k_3} \quad = \; \Symbol{6j} \quad .
	\label{g:6j}
\end{equation}
(To obtain \Eq{g:6j}, contract both sides of \Eq{g:13-in-basis-12} with an intertwiner of the form \eqref{g:iota4}, keeping in mind \Eq{iota4-norm}.) As seen from \Eq{g:6j}, the $6j$-symbol represents a complete contraction of four $3j$-symbols. The $6j$-symbol vanishes unless the triangular conditions are satisfied by the four triples of spins indicated by
\begin{equation}
	\sixj{\circ}{\circ}{\circ}{}{}{} \qquad \sixj{\circ}{}{}{}{\circ}{\circ} \qquad \sixj{}{\circ}{}{\circ}{}{\circ} \qquad \sixj{}{}{\circ}{\circ}{\circ}{}{}.
	\label{}
\end{equation}
The symmetry properties of the $6j$-symbol include invariance under any permutation of its columns,
\begin{equation}
	\sixj{j_1}{j_2}{j_3}{k_1}{k_2}{k_3} = \sixj{j_1}{j_3}{j_2}{k_1}{k_3}{k_2} = \sixj{j_2}{j_3}{j_1}{k_2}{k_3}{k_1}, \quad \text{etc.}
	\label{}
\end{equation}
as well as invariance under an interchange of the upper and lower spins in any two columns,
\begin{equation}
	\sixj{j_1}{j_2}{j_3}{k_1}{k_2}{k_3} = \sixj{j_1}{k_2}{k_3}{k_1}{j_2}{j_3} = \sixj{k_1}{k_2}{j_3}{j_1}{j_2}{k_3}, \quad \text{etc.}
	\label{}
\end{equation}
Invariant contractions of greater numbers of $3j$-symbols give rise to higher Wigner $nj$-symbols.  For instance, the change of basis between two bases of five-valent intertwiners is encoded in the $9j$-symbol, which is a complete contraction of six $3j$-symbols. These symbols and their properties are described in more detail in references such as \cite{brink1968angular, Varshalovich:1988ye, Yutsis:1962bk}.

\subsection{Example: The gauge invariant projector}

To give an example of using the elements introduced in this chapter in a concrete calculation, let us consider the integral
\begin{equation}
	I^{(j_1\cdots j_N)} \equiv \int dg\,D^{(j_1)}(g)\cdots D^{(j_N)}(g).
	\label{integral}
\end{equation}
We would like to show that, seen as an operator on the space ${\cal H}_{j_1} \otimes \cdots \otimes {\cal H}_{j_N}$, the object \eqref{integral} represents the orthogonal projector onto the gauge invariant subspace ${\rm Inv}\,\bigl({\cal H}_{j_1} \otimes \cdots \otimes {\cal H}_{j_N}\bigr)$.

The product of $N$ representation matrices is described graphically by the expression
\begin{equation}
	\Symbol{D_product} \quad . 
	\label{}
\end{equation}
Now the strategy is to repeatedly use \Eq{g:DD} to couple the two leftmost representation matrices. Proceeding in this way until there are only two matrices remaining, we arrive at
\begin{equation}
	\sum_{k_1\cdots k_{N-2}} d_{k_1}\cdots d_{k_{N-2}}\quad\Symbol{all_coupled} \quad . 
	\label{}
\end{equation}
At this point we may compute the integral over the group using the orthogonality theorem \eqref{g:intDD} for the Wigner matrices. We then find that the integral \eqref{integral} is equal to
\begin{equation}
	\sum_{k_1\cdots k_{N-3}} d_{k_1}\cdots d_{k_{N-3}}\quad \Symbol{projector} \quad . 
	\label{}
\end{equation}
Here we can recognize the graphical representation of the $N$-valent intertwiners \eqref{g:iotaN}, which span the intertwiner space ${\rm Inv}\,\bigl({\cal H}_{j_1} \otimes \cdots \otimes {\cal H}_{j_N}\bigr)$. Hence, introducing the label $\vec k$ to denote the internal spins $k_1, \dots, k_{N-3}$, we have shown that
\begin{equation}
	I^{(j_1\cdots j_N)}{}\updown{m_1\cdots m_N}{n_1\cdots n_N} = \sum_{k_1\cdots k_{N-3}} d_{k_1}\cdots d_{k_{N-3}}\bigl(\iota^{(j_1\cdots j_N)}_{\vec k}\bigr)^{m_1\cdots m_N}\bigl(\iota^{(j_1\cdots j_N)}_{\vec k}\bigr)_{n_1\cdots n_N}.
	\label{}
\end{equation}
Using the normalized intertwiners $\nu^{(j_1\cdots j_N)}_{\vec k} = \sqrt{d_{k_1}\cdots d_{k_{N-3}}}\,\iota^{(j_1\cdots j_N)}_{\vec k}$ we may write
\begin{equation}
	I^{(j_1\cdots j_N)} = \sum_{\vec k}\,\bket{\nu^{(j_1\cdots j_N)}_{\vec k}}\bbra{\nu^{(j_1\cdots j_N)}_{\vec k}}, 
	\label{projector}
\end{equation}
which establishes the statement we were looking to prove, since \Eq{projector} is nothing but the standard way of expressing a projection operator in terms of an orthonormal basis of the subspace onto which the projection is taken.

\section{Calculating with graphical diagrams}
\label{sec:calculating}

\subsection{The fundamental theorem of graphical calculus}

In section \ref{sec:elements} we have introduced the basic ingredients of graphical calculus, consisting of the graphical representation of objects of $SU(2)$ recoupling theory, together with rules such as \Eqs{g:eps_nm}--\eqref{g:epseps2} and \eqref{g:3j-minus}--\eqref{g:3j-arrows} which show how the arrows and signs in graphical diagrams can be manipulated. However, the most powerful element of the graphical method emerges, perhaps surprisingly, from the simple fact that an $N$-valent invariant tensor can be expanded with respect to a given basis of the corresponding $N$-valent intertwiner space. Due to their indispensable role in graphical calculations, we will refer to the identities derived from this observation as the \idx{fundamental theorem of graphical calculus} (although this is not a standard terminology in the literature of the subject). To express these identities in graphical form, we will introduce a notation where an $N$-valent invariant tensor is represented by a block with $N$ lines attached.

A tensor carrying a single index can be invariant only if the index belongs to the trivial representation. Thus, identifying a line having $j=0$ with no line, we have
\begin{equation}
	\Symbol{1block} \quad = \quad \delta_{j,0}\quad\Symbol{1block-noline} \quad . 
	\label{thm1}
\end{equation}
A two-valent invariant tensor $T_{mn}$ must be proportional to $\epsilon^{(j)}_{mn}$, which is the only invariant tensor having two lower indices. In particular, $T_{mn}$ cannot be invariant unless both of its indices belong to the same representation. The coefficient of proportionality can be deduced by contracting both sides of the equation with the epsilon tensor. In this way we obtain the graphical identity
\begin{equation}
	\Symbol{2block} \quad = \quad \delta_{jj'}\frac{1}{d_j}\quad\Symbol{2block-exp} \quad . \vspace{4pt}
	\label{thm2}
\end{equation}
The three-valent intertwiner space ${\rm Inv}\,\bigl({\cal H}_{j_1}\otimes{\cal H}_{j_2}\otimes{\cal H}_{j_3}\bigr)$ is also one-dimensional and is spanned by the $3j$-symbol \eqref{g:iota3}. A three-valent invariant tensor must therefore be proportional to the $3j$-symbol, with the coefficient of proportionality determined by contracting both sides of the equation with the $3j$-symbol:
\begin{equation}
	\Symbol{3block} \quad = \quad \Symbol{3block-exp} \quad . 
	\label{thm3}
\end{equation}
A four-valent invariant tensor can be expanded using any basis of the four-valent intertwiner space ${\rm Inv}\,\bigl({\cal H}_{j_1}\otimes\cdots\otimes{\cal H}_{j_4}\bigr)$. The dimension of this space is in general higher than $1$, so several possible choices of basis are available. If we take the basis given by \Eq{g:iota4} and keep in mind the normalization of the basis intertwiners, we can write
\vspace{-4pt}
\begin{equation}
	\Symbol{4block} \quad = \quad \sum_x d_x\quad\Symbol{4block-exp} \quad . 
	\label{thm4}
\end{equation}
Different versions of the identity \eqref{thm4} can be derived by using different bases of the four-valent intertwiner space. For example, choosing the basis \eqref{g:iota4-13}, we obtain
\begin{equation}
	\Symbol{4block} \quad = \quad \sum_x d_x\quad\Symbol{4block-exp-13} \quad . 
	\label{thm4-13}
\end{equation}
Invariant tensors of valence higher than four can be expanded in a way completely analogous to \Eqs{thm4} and \eqref{thm4-13}, using any basis of the appropriate intertwiner space.

An important use of the fundamental theorem of graphical calculus is to simplify graphical diagrams which carry no external, uncontracted lines, and which therefore represent invariant contractions of $3j$-symbols\footnote{If the invariance of a given closed diagram constructed out of $3j$-symbols and epsilon tensors is not immediately apparent, it can be checked by the following criterion: The diagram is invariant if and only if \Eqs{g:epseps2} and \eqref{g:3j-arrows} can be used to transform it to a form where each line carries exactly one arrow.}. Suppose that an invariant diagram contains a subdiagram which itself represents an $N$-valent invariant tensor, and that the entire diagram can be divided into two disconnected pieces by cutting the $N$ external lines of the subdiagram. In such a case the diagram can be simplified (or at least transformed into a different form) by applying the fundamental theorem to the $N$-valent intertwiner corresponding to the subdiagram. For diagrams which can be separated into two pieces by cutting one, two, three or four lines, this procedure gives rise to the following graphical identities:
{
\allowdisplaybreaks
\begin{align}
	\Symbol{1blocks} \quad &= \quad \delta_{j,0}\quad\Symbol{1blocks-cut} \quad ,  \label{thm1'} \\[8pt]
	\Symbol{2blocks} \quad &= \quad \delta_{jj'}\frac{1}{d_j}\quad\Symbol{2blocks-cut} \quad ,  \label{thm2'} \\[8pt]
	\Symbol{3blocks} \quad &= \quad \Symbol{3blocks-cut} \quad ,  \label{thm3'} \\[8pt]
	\Symbol{4blocks} \quad &= \quad \sum_x d_x\quad\Symbol{4blocks-cut} \quad .  \label{thm4'}
\end{align}
As before, different versions of \Eq{thm4'} can be obtained by selecting different bases on the four-valent intertwiner space. Equation \eqref{thm4'} also generalizes in a straightforward way to the case of cutting a diagram across more than four lines.
}

\subsection{Example: Biedenharn--Elliot identity}

An instructive example of using the fundamental theorem of graphical calculus is provided by a graphical proof of the Biedenharn--Elliot identity
\begin{align}
	&\sixj{j_1}{j_2}{j_3}{k_1}{k_2}{k_3}\sixj{j_1}{j_2}{j_3}{l_1}{l_2}{l_3} \notag \\
	&= \sum_x d_x(-1)^{j_1+j_2+j_3+k_1+k_2+k_3+l_1+l_2+l_3+x}\sixj{j_1}{k_2}{k_3}{x}{l_3}{l_2}
	\sixj{k_1}{j_2}{k_3}{l_3}{x}{l_1}\sixj{k_1}{k_2}{j_3}{l_2}{l_1}{x}.
	\label{biedenharn-elliot}
\end{align}
In loop quantum gravity the Biedenharn--Elliot identity is relevant to the Ponzano--Regge model \cite{Barrett:2009,Ponzano:1968}, a spin foam quantization of three-dimensional Euclidean gravity, where it can be used to show that the partition function of the model is invariant under the so-called 2--3 Pachner move.

We begin from the left-hand side of \Eq{biedenharn-elliot}, using \Eq{g:6j} to express the $6j$-symbols in graphical form. For the second $6j$-symbol we invoke the following theorem about closed diagrams: The value of a diagram representing an invariant contraction of $3j$-symbols is preserved if all the signs and all the arrows in the diagram are simultaneously reversed\footnote{The proof of this statement is straightforward. After the invariant diagram has been put into a form where each line carries an arrow, reversing all the arrows produces the factor $(-1)^{2J}$, where $J$ is the sum of all the spins in the diagram. Reversing the sign at a node with spins $j_1$, $j_2$ and $j_3$ multiplies the diagram by $(-1)^{j_1+j_2+j_3}$, so reversing all the signs also produces the factor $(-1)^{2J}$, since every line is connected to exactly two nodes. The total factor arising from the process is therefore $(-1)^{4J} = +1$.}. Thus, we have
\begin{equation}
	\sixj{j_1}{j_2}{j_3}{k_1}{k_2}{k_3}\sixj{j_1}{j_2}{j_3}{l_1}{l_2}{l_3} \quad = \quad \Symbol{6j6j} \quad . 
	\label{BE-lhs}
\end{equation}
Now we can see that the structure of the graphical diagram matches with the pattern on the right-hand side of \Eq{thm3'}. We may therefore use the fundamental theorem ``in reverse'' to join the two $6j$-symbols into a single connected diagram as follows:
\begin{equation}
	\Symbol{6j6j} \quad = \quad \Symbol{6j6j_connected} \quad . 
	\label{BE-cut}
\end{equation}
(After forming the connected diagram we have noted that, as shown by \Eq{g:epseps2}, the two oppositely oriented arrows cancel on the lines carrying spins $j_1$ and $j_3$.)

The next step will be to cut the diagram on the right-hand side across four lines, using the fundamental theorem in the form \eqref{thm4'}. Before doing so, we introduce a triple of outwardly oriented arrows, as shown by \Eq{g:3j-arrows}, on each of the three nodes carrying a $+$ sign in order to ensure that each piece resulting from the cut will represent a proper invariant contraction. We then obtain {
\allowdisplaybreaks
\begin{align}
	&\Symbol{6j6j_connected_arrows} \quad = \quad \sum_x d_x\;\Symbol{6j_j1}\quad\Symbol{6j6j_after_cut} \notag \\
	&= \quad \sum d_x \; \Symbol{6j_j1} \Symbol{6j_j2} \quad \Symbol{6j_j3},
	\label{6j6j_cut}
\end{align}
where we have first applied \Eq{thm4'} to cut the diagram across the lines carrying spins $k_2$, $k_3$, $l_2$ and $l_3$, and then used the fundamental theorem again to split the new diagram in two pieces by cutting the three lines that connect the inner part of the diagram to the outer part. Now each diagram on the second line of \Eq{6j6j_cut} can be recognized as a $6j$-symbol by comparing with the reference diagram \eqref{g:6j}, adjusting the signs and arrows as needed. (It is useful to remember that simultaneously reversing all the signs and all the arrows does not change the value of a diagram.) For instance, the first diagram is equal to
\vspace{-4pt}
\begin{equation}
	(-1)^{j_1+k_2+k_3}(-1)^{2k_2}\sixj{j_1}{k_2}{k_3}{x}{l_3}{l_2}. 
	\label{}
\end{equation}
To obtain the identity \eqref{biedenharn-elliot}, it only remains to carefully keep track of the factors of $(-1)$, noting that the triangular conditions satisfied by the triples $(k_1, l_1, x)$ and $(k_2, l_2, x)$ imply that $(-1)^{2k_1+2l_1+2x} = 1$ and $(-1)^{2k_2+2l_2+2x} = 1$.
}

\section{The graphical method in loop quantum gravity}
\label{sec:LQG}

\subsection{Kinematical states and elementary operators}

The kinematical Hilbert space of loop quantum gravity is spanned by the (generalized) spin network states. A generalized \idx{spin network state} is labeled by an oriented graph $\gamma$ together with a set of spins $\vec j = \{j_1, \dots, j_{N_e}\}$ associated to the edges of the graph, and a set of $SU(2)$ tensors (generalized intertwiners) $\vec\iota = \{\iota_1, \dots, \iota_{N_v}\}$ associated to the vertices of the graph (here $N_e$ and $N_v$ denote the number of edges and vertices of the graph). The state is a function of $N_e$ $SU(2)$ group elements, and is defined by
\begin{equation}
	T_{\gamma,\vec{j},\vec{\iota}}(h_{e_1}, \dots, h_{e_N}) = \bigotimes_{v\in V(\gamma)} \iota_v\cdot \bigotimes_{e\in E(\gamma)} D^{(j_e)}(h_e).
	\label{spin-network-state}
\end{equation}
The index structure of the tensor $\iota_v$ at a given vertex is adapted to the structure edges incident to the vertex: If the vertex contains $M$ edges oriented outwards and labeled by spins $j_1, \dots, j_M$ and $N-M$ edges oriented inwards and labeled by spins $j_{M+1}, \dots, j_N$, the tensor $\iota_v$ carries $M$ upper indices in the representations $j_1, \dots, j_M$ and $N-M$ lower indices in the representations $j_{M+1}, \dots, j_N$. The dot in \Eq{spin-network-state} denotes a complete contraction of magnetic indices between the representation matrices and the generalized intertwiners. By restricting the set of tensors $\iota_v$ at each vertex to be the invariant tensors (proper intertwiners) described in section \ref{sec:intertwiners}, one obtains the proper spin network states, which are invariant under local $SU(2)$ gauge transformations -- see \Eq{holonomy-gauge} -- and which form  a basis on the gauge-invariant Hilbert space of loop quantum gravity.

The group elements in \Eq{spin-network-state} originate classically from holonomies of the Ashtekar connection along curves in the spatial manifold $\Sigma$. Given an edge $e : [0,1] \to \Sigma$, the holonomy of the Ashtekar connection $A_a^i$ along the edge is
\begin{align}
	h_e[A] &= {\cal P}\exp\biggl(-\int_e A\biggr) \notag \\
	&= \Id + \sum_{n=1}^\infty (-1)^n\,\int_0^1 dt_1\int_0^{t_1} dt_2\,\dots \int_0^{t_{n-1}}dt_n\,A\bigl(e(t_1)\bigr)\cdots A\bigl(e(t_n)\bigr),
	\label{}
\end{align}
where $A\bigl(e(t)\bigr) \equiv \dot e^a(t)A_a^i\bigl(e(t)\bigr)\tau_i$, with $\dot e^a(t)$ being the tangent vector of $e$, and ${\cal P}\exp$ denotes the path ordered exponential, with the largest path parameter ordered to the left. The holonomy satisfies certain properties which reflect its geometric interpretation as a parallel propagator. Perhaps the most important of these is the identity
\begin{equation}
	h_{e_2\circ e_1} = h_{e_2}h_{e_1},
	\label{h2h1}
\end{equation}
where $e_1$ and $e_2$ are two edges such that the endpoint of $e_1$ coincides with the beginning point of $e_2$, and $e_2\circ e_1$ is the combined path formed by $e_1$ followed by $e_2$.

Under a local $SU(2)$ gauge transformation described by a gauge function $g(x) \in SU(2)$, the connection behaves as $A_a \to (A^g)_a \equiv gA_ag^{-1} + g\partial_ag^{-1}$. This implies the corresponding transformation law of the holonomy as
\begin{equation}
h_e[A^g] = g\bigl(f(e)\bigr)h_e[A]g^{-1}\bigl(b(e)\bigr),
\label{holonomy-gauge}
\end{equation}
where $b(e)$ and $f(e)$ denote the beginning and final points of $e$. Equation \eqref{holonomy-gauge} suggests that the indices $m$ and $n$ of the representation matrix $\DD{j}{m}{n}{h_e[A]}$ are associated respectively with the final and beginning points of the edge $e$. Our convention for the graphical representation of the Wigner matrices, given by \Eq{g:D}, has been chosen accordingly so that the direction of the triangle is consistent with the orientation of the edge in the holonomy $h_e[A]$.

The elementary operators of loop quantum gravity are the holonomy and flux operators. The \idx{holonomy operator} acts as a multiplicative operator:
\begin{equation}
	\widehat{\DD{j}{m}{n}{h_e}} T_{\gamma,\vec{j},\vec{\iota}}(h_{e_1}, \dots, h_{e_N}) = \DD{j}{m}{n}{h_e}T_{\gamma,\vec{j},\vec{\iota}}(h_{e_1}, \dots, h_{e_N}).
	\label{}
\end{equation}
If the edge $e$ coincides with one of the edges $e_1, \dots, e_N$, the action of the holonomy operator on a spin network state involves essentially the decomposition of the tensor product of two representations, which is given by the Clebsch--Gordan series \eqref{DD}.

The flux operator arises as a quantization of the classical variable representing the flux of the densitized triad $E^a_i$ through a 2-dimensional surface. The action of the flux operator on a spin network state can be expressed as a linear combination of the \idx{left-} and right-\idx{invariant vector fields} of $SU(2)$. The left- and \idx{right-invariant vector fields} act according to the definitions
\begin{align}
	\hat L_i^{(e)}T_{\gamma, \vec j, \vec\iota}(h_{e_1}, \dots, h_{e_N}) &= i\frac{d}{dt}\bigg|_{t=0}T_{\gamma, \vec j, \vec\iota}(h_{e_1}, \dots, h_ee^{t\tau_i}, \dots h_{e_N}), \label{L_i} \\
	\hat R_i^{(e)}T_{\gamma, \vec j, \vec\iota}(h_{e_1}, \dots, h_{e_N}) &= -i\frac{d}{dt}\bigg|_{t=0}T_{\gamma, \vec j, \vec\iota}(h_{e_1}, \dots, e^{t\tau_i}h_e, \dots h_{e_N}), \label{R_i}
\end{align}
where the superscripts indicate that the operators act on the argument $h_e$ of the function $T_{\gamma, \vec j, \vec\iota}(h_{e_1}, \dots, h_{e_N})$. If these operators are applied on the Wigner matrices themselves, it is immediate to see that their action is given by
\begin{align}
	\hat L_i^{(e)}D^{(j)}(h_e) &= iD^{(j)}(h_e)\tau_i^{(j)}, \label{L*D} \\
	\hat R_i^{(e)}D^{(j)}(h_e) &= -i\tau_i^{(j)}D^{(j)}(h_e). \label{R*D}
\end{align}

To apply the methods of graphical calculus for calculations in loop quantum gravity, the general strategy is as follows. Firstly, one expresses the spin network states in a graphical form. Secondly, one derives graphically the action of the two elementary operators on the spin network states. After this has been done, the action of any well-defined operator constructed out of the elementary operators can be derived by applying the graphical formalism introduced in sections \ref{sec:elements} and \ref{sec:calculating}.

The spin network states defined by \Eq{spin-network-state} involve two ingredients: The representation matrices $D^{(j)}(h_e)$ and the intertwiners $\iota_v$. The graphical representation of the Wigner matrices is given by \Eq{g:D}, while the graphical representation of intertwiners has been discussed in section \ref{sec:intertwiners}. Let $T_{\gamma, \vec j, \vec\iota}^v$ denote the part of a spin network state directly associated to the vertex $v$. Assume that the vertex $v$ contains $N$ edges $e_1, \dots, e_N$, the first $M$ of which are oriented outwards and the remaining $N-M$ are oriented inwards. If the intertwiner at $v$ is a tensor of the form \eqref{g:iotaN}, the function $T_{\gamma, \vec j, \vec\iota}^v$ is represented graphically by the expression
\begin{equation}
	T_{\gamma, \vec j, \vec\iota}^v(h_{e_1}, \dots, h_{e_N}) \; = \; \Symbol{node} \quad .
	\label{spin-network-vertex}
\end{equation}
Note that, for each edge oriented outwards, the corresponding index of the intertwiner has been raised using the epsilon tensor.

Now let us consider the graphical calculation of the action of the two elementary operators on spin network states. The action of the holonomy operator is encoded in the Clebsch--Gordan series, which is given in algebraic form by \Eq{DD} and in graphical form by \Eq{g:DD}. Applying the holonomy operator to the state \eqref{spin-network-vertex}, and using \Eq{g:DD} to evaluate the resulting action, we find that the action of the holonomy operator can be represented in graphical form as
\begin{align}
	&\widehat{\DD{j}{m}{n}{h_e}} T_{\gamma, \vec j, \vec\iota}^v(h_{e_1}, \dots, h_e, \dots, h_{e_N}) \notag \\
	&= \; \sum_k d_k \; \Symbol{D-action} \quad . 
	\label{D*T}
\end{align}

Consider then the left- and right-invariant vector fields. We want to obtain a graphical expression for the action of the operators $\hat L_i^{(e)}$ and $\hat R_i^{(e)}$ on the function $T_{\gamma, \vec j, \vec\iota}^v(h_{e_1}, \dots, h_{e_N})$. As the action of the left- and right-invariant vector fields on a representation matrix is given by \Eqs{L*D} and \eqref{R*D}, it suffices to recall the graphical representation of the $SU(2)$ generators $\tau_i^{(j)}$ from \Eq{g:tau} to cast the action of these operators into graphical form. Applying this to the state represented by the diagram \eqref{spin-network-vertex}, we establish that the action of the left- and right-invariant vector fields on a spin network vertex is given by the graphical expression
\begin{equation}
	\hat L_i^{(e)} T_{\gamma, \vec j, \vec\iota}^v \; = \; -W_{j_e} \; \Symbol{L-action} \quad ,
	\label{g:L}
\end{equation}
and
\begin{equation}
	\hat R_i^{(e)} T_{\gamma, \vec j, \vec\iota}^v \; = \; W_{j_e} \; \Symbol{R-action} \quad .
	\label{g:R}
\end{equation}
Note that there is a certain sense of consistency between the invariant vector fields and the orientation of the edge on which they act. If a left-invariant vector field is applied to an outgoing edge, or a right-invariant vector field to an incoming edge, the action of the operator is localized to the vertex $v$ and can be interpreted essentially as an action on just the intertwiner $\iota_v$.

Equations \eqref{D*T}--\eqref{g:R} provide the graphical representation of the elementary operators of loop quantum gravity. The action of any well-defined operator on a spin network state can now be computed graphically by applying these equations together with the basic rules of graphical calculus presented in sections \ref{sec:elements} and \ref{sec:calculating}.

\subsection{Example: Matrix elements of the Hamiltonian constraint}

The task of computing the matrix elements of the Hamiltonian constraint, which is the operator governing the dynamics in the canonical formulation of loop quantum gravity, is a typical example of a calculation which can be performed quite efficiently using the graphical methods presented in this chapter. As an example, let us take a look at a version of the Hamiltonian introduced in \cite{Yang:2015zda}. The operator is a slight variation of Thiemann's well-known construction \cite{Thiemann:1996aw}, and it has been studied \eg in \cite{Yang:2021den} to investigate the consistency between the canonical and covariant formulations of the dynamics in loop quantum gravity.

The operator corresponds to the Euclidean part of the Hamiltonian constraint, and restricted to a given node $v$ of a spin network state, it takes the form
\begin{equation}
	\hat H^{\rm E}_v = \sum_{\text{$e_I$, $e_J$ at $v$}} \epsilon^{ijk}\,{\rm Tr}\,\Bigl(\tau_k^{(l)}\widehat{D^{(l)}(h_{\alpha_{IJ}})}\Bigr)\hat J_i^{(e_I)}\hat J_j^{(e_J)}\widehat{V_v^{-1}}.
	\label{H_E}
\end{equation}
Here $h_{\alpha_{IJ}}$ is the holonomy around the closed triangular loop $\alpha_{IJ} = s_I^{-1}\circ a_{IJ}\circ s_J$, where $s_I$ and $s_J$ are short segments of the edges $e_I$ and $e_J$ starting from $v$, and $a_{IJ}$ is an arc connecting the endpoint of $s_J$ to the endpoint of $s_I$. Moreover, each $\hat J_i^{(e)}$ denotes either a left- or a right-invariant vector field, according to whether the corresponding edge is oriented outwards or inwards at the vertex -- see the remark below \Eq{g:R}, and $\widehat{V_v^{-1}}$ is a regularized inverse volume operator (see \eg \cite{Bianchi:2008es, Yang:2021den}).
For the purposes of our example, we will ignore the inverse volume operator, whose action is not accessible by purely graphical means, and focus on the remaining part of operator \eqref{H_E}:
\begin{equation}
	\hat h^{\rm E}_{v, e_I, e_J} = \epsilon^{ijk}\,{\rm Tr}\,\Bigl(\tau_k^{(l)}\widehat{D^{(l)}(h_{\alpha_{IJ}})}\Bigr)\hat J_i^{(e_I)}\hat J_j^{(e_J)}.
	\label{h_E}
\end{equation}
Our goal is to calculate the action of the operator $\hat h^{\rm E}_{v, e_1, e_2}$ on a four-valent spin network vertex. For simplicity, assume that all the edges are oriented outwards from the vertex. Then the relevant part of the spin network function has the form\footnote{
	In the name of readability, we commit a slight abuse of notation by using the edges $e_I$ instead of the holonomies $h_{e_I}$ to label the triangles in the diagram representing the spin network state.
}
\begin{equation}
	\DD{j_1}{m_1}{n_1}{h_{e_1}} \cdots \DD{j_4}{m_4}{n_4}{h_{e_4}}\bigl(\iota^{(j_1\cdots j_4)}_k\bigr)^{n_1\cdots n_4} \quad = \quad \Symbol{4node-e},
	\label{4node}
\end{equation}
where we have chosen a basis of intertwiners where the edges $e_1$ and $e_2$, on which the operator $\hat h^{\rm E}_{v, e_1, e_2}$ will act, are coupled to the internal spin of the intertwiner.

Each $\hat J_i^{(e)}$ in \Eq{h_E} will now be a left-invariant vector field. Using \Eq{g:L} to evaluate the action of these operators on the state \eqref{4node}, and appending the diagram
\begin{equation}
	\epsilon^{ijk}\Tau{l}{k}{m}{n} \; = \; -\sqrt 6 W_l\; \Symbol{eps_tau}
	\label{}
\end{equation}
to the resulting expression, we obtain
\begin{align}
	\epsilon^{ijk}\Tau{l}{k}{m}{n}\hat J_i^{(e_1)}\hat J_j^{(e_2)}\bket{T^v_{\vec j, k}} \; = \; -\sqrt 6 W_{j_1}W_{j_2}W_l\; \Symbol{JJ_action},
	\label{}
\end{align} 
where $\bket{T^v_{\vec j, k}}$ denotes the state \eqref{4node}. We must then bring in the operator $\widehat{\DD{l}{n}{m}{h_{\alpha_{12}}}}$. Using the multiplicative property \eqref{h2h1} of the holonomy, we can break this down as
\begin{equation}
	\widehat{\DD{l}{n}{m}{h_{\alpha_{12}}}} = \widehat{\DD{l}{n}{n'}{h_{s_1}^{-1}}} \widehat{\DD{l}{n'}{m'}{h_{a_{12}}}} \widehat{\DD{l}{m'}{m}{h_{s_2}}}.
	\label{}
\end{equation}
The action of the rightmost holonomy operator on the holonomy $D^{(j_2)}(h_{e_2})$ can be derived by splitting the edge $e_2$ into the segment $s_2$ and the remainder $e_2'$ -- recall \Eq{h2h1} -- and applying \Eq{D*T}. After slightly adjusting the direction of the arrows, we find
\begin{align}
	\widehat{\DD{l}{m'}{m}{h_{s_2}}} \; \Symbol{h_e2} \quad &= \quad \widehat{\DD{l}{m'}{m}{h_{s_2}}} \; \Symbol{h_e2_split} \notag \\
	&= \quad \sum_{l_2} d_{l_2}\,(-1)^{2l_2}\;\Symbol{h_e2_coupled}.
	\label{}
\end{align}
The operator $\widehat{\DD{l}{n}{n'}{h_{s_1}^{-1}}}$ can be treated in the same way, using \Eq{D-inv} for the matrix elements of the inverse matrix and taking into account the arrows arising from the epsilon tensors. Finally, the holonomy
\begin{equation}
	\DD{l}{n'}{m'}{h_{a_{12}}} \; = \; \Symbol{h_a12}
	\label{}
\end{equation}
is attached to the diagram. In the end we have
\begin{align}
	&\hat h^{\rm E}_{v, e_1, e_2}\bket{T^v_{\vec j, k}} \; = \; -\sqrt 6 W_{j_1}W_{j_2}W_l \notag \\
	&\times \sum_{l_1, l_2} d_{l_1}d_{l_2}(-1)^{2l_1+2l_2} \Symbol{h12_action}.
	\label{h12_action}
\end{align}
The next step is to express the result in terms of known quantities such as $6j$-symbols with the help of the fundamental theorem. We begin by applying \Eq{thm3} to the three-valent invariant tensor within the above diagram, thus finding that the diagram is equal to
\begin{equation}
	f(j_1, j_2, k, l, l_1, l_2) \quad \Symbol{h12_action_final},
	\label{h12_state}
\end{equation}
where
\begin{equation}
	f(j_1, j_2, k, l, l_1, l_2) \; = \quad \Symbol{12j}.
	\label{12j}
\end{equation}
This diagram can be recognized as the so-called $12j$-symbol of the second kind (see \eg \cite{Varshalovich:1988ye, Yutsis:1962bk}), but we may equally well continue to use the fundamental theorem to break it down. Clearly the diagram cannot be split into two disconnected pieces by cutting only three lines. If we cut the four lines connecting the central part of the diagram to the two outer nodes, as indicated by \Eq{thm4'}, we see that
\begin{equation}
	f(j_1, j_2, k, l, l_1, l_2) \; = \; \sum_x d_x \; \Symbol{12j_piece1} \quad \Symbol{12j_piece2}
	\label{}
\end{equation}
Here the first piece is immediately recognizable as a $6j$-symbol, while the second piece can be reduced to a product of three $6j$-symbols by making two horizontal cuts. After simplifying the factors of $(-1)$, we obtain the final result of our calculation in the form
\begin{align}
	\hat h^{\rm E}_{v, e_1, e_2} \bket{T^v_{\vec j, k}} ={} &\sqrt 6 W_{j_1}W_{j_2}W_l \sum_{l_1, l_2,x} d_{l_1}d_{l_2}d_x\,(-1)^{k+l_1-l_2} \notag \\
	&\times \sixj{j_1}{j_2}{k}{l_2}{l_1}{x}\sixj{j_1}{j_1}{1}{x}{l}{l_1}\sixj{j_2}{j_2}{1}{x}{l}{l_2}\sixj{1}{1}{1}{x}{l}{l}\bket{\widetilde T^v_{\vec j, \vec l, k}},
	\label{h12_final}
\end{align}
where $\bket{\widetilde T^v_{\vec j, \vec l, k}}$ is the state defined by the diagram in \Eq{h12_state}.

\section{Summary}
\label{sec:summary}

In this chapter we have presented an introduction to the graphical calculus of $SU(2)$, which constitutes a highly efficient technique for calculations involving $SU(2)$ recoupling theory. The graphical formalism, which has been originally introduced in the literature as a method for dealing with calculations in the quantum theory of angular momentum, is built out of two key ingredients: (1) A graphical notation consisting of diagrammatic representations of the elementary objects of $SU(2)$ recoupling theory, such as Clebsch--Gordan coefficients, Wigner $nj$-symbols and Wigner matrices; and (2) The basic properties satisfied by these graphical diagrams, which follow from the properties of the corresponding non-graphical objects, and which promote the graphical notation into a powerful diagrammatic calculus. In particular, in Sec.~\ref{sec:calculating} we introduced a set of graphical identities which we named the fundamental theorem of graphical calculus; these identities provide an essential tool for simplifying complicated graphical diagrams.

In the graphical approach, any given algebraic expression in $SU(2)$ recoupling theory is represented in a definite and unambiguous way by a corresponding graphical formula. Calculations can then be performed graphically by following a set of straightforward rules for manipulating graphical diagrams. The resulting transformations of the graph correspond uniquely to algebraic manipulations of the corresponding non-graphical expressions, so any calculation performed by the graphical method can also be performed by conventional algebraic techniques. However, the graphical form of the calculation is usually considerably more concise, efficient and visually easier to follow.

The graphical calculus of $SU(2)$ provides a very effective practical tool for calculations involving the spin network states of loop quantum gravity, which have the structure of $SU(2)$ representation matrices, associated with the edges of a graph, contracted with intertwiners, or invariant tensors of $SU(2)$, associated with the nodes of the graph. By applying the basic definitions of the graphical formalism, the action of the elementary operators of loop quantum gravity can be cast into a graphical form. Given an operator constructed out of holonomy and flux operators, the action of the operator on spin network states can then be systematically computed by applying the rules of graphical calculus. As a concrete example, we considered the calculation of the matrix elements of a particular version of the Hamiltonian constraint operator in the spin network basis. Many more examples of the use of graphical calculus in loop quantum gravity can be found in the articles cited in the bibliography, where graphical techniques have been successfully applied to a variety of physically relevant and technically quite non-trivial calculations.

\section*{Acknowledgments}

I. M. acknowledges support of the National Science Centre, Poland through grants no. 2018/30/Q/ST2/00811 and 2022/44/C/ST2/00023. J. Y. is supported in part by NSFC Grants No. 12165005 and No. 11961131013.

\end{document}